\documentclass[prb,aps,twocolumn,showpacs,groupedaddress]{revtex4}

\usepackage{graphicx}
\usepackage{color}
\usepackage{pstricks}
\usepackage{amsmath}
\usepackage{dsfont}

\def\unit#1{\ \mathrm{#1}}

\def\vek#1{\mathbf{#1}}
\def\arcsinh{\mathrm{arcsinh}\, }
\def\arcsin{\mathrm{arcsin}\, }

\begin{document}

\title{Microscopic mechanism of the non-crystalline anisotropic
  magnetoresistance \par in (Ga,Mn)As}

\author{Karel V\'yborn\'y$^{1}$, Jan Ku\v cera$^{1}$, 
Jairo Sinova$^{2,1}$, A.W. Rushforth$^{3}$, B.L.~Gallagher$^{3}$, and 
T. Jungwirth$^{1,3}$}

\address{$^{1}$\hbox{Institute of Physics of the ASCR,  v. v. i., 
Cukrovarnick\'a 10, Praha 6, CZ--16253, Czech Republic}}
\address{$^{2}$Department of Physics, Texas A\&M University, College Station,
  TX 77843-4242, USA}
\address{$^{3}$School of Physics and Astronomy, University of Nottingham,
  Nottingham NG7 2RD, United Kingdom}

\date{Apr 25th, 2009}

\begin{abstract}
  Starting with a microscopic model based on the Kohn-Luttinger Hamiltonian
  and kinetic $p$-$d$ exchange combined with Boltzmann formula for
  conductivity we identify the scattering from magnetic Mn combined with the
  strong spin-orbit interaction of the GaAs valence band as the dominant
  mechanism of the anisotropic magnetoresistance (AMR) in (Ga,Mn)As. This fact
  allows to construct a simple analytical model of the AMR consisting of two
  heavy-hole bands whose charge carriers are scattered on the impurity
  potential of the Mn atoms. The model predicts the correct sign of the AMR
  (resistivity parallel to magnetization is smaller than perpendicular to
  magnetization) and identifies its origin arising from the destructive
  interference between electric and magnetic part of the scattering potential
  of magnetic ionized Mn acceptors when the carriers move parallel to the
  magnetization.
\end{abstract}


\maketitle

\section{Introduction}

Although it has been known for 150 years that electric resistance of a
magnetic metal depends on the direction of magnetization,\cite{Thomson:1857_a}
the origin of such dependence is often explained only vaguely as an interplay
of spin-orbit interaction (SOI) and magnetization.  Conceptual questions
around this phenomenon, the anisotropic magnetoresistance (AMR), remain open
and relate to the quest for its detailed mechanism, its sign and specific
magnitude.  Our ability to control the AMR by material design, with potential
impact on new electronic devices,\cite{Chappert:2007_a} would be improved if
we had answers more specific but nevertheless still more universal than
``black-box-like'' modelling of AMR in each and every deemable system.

Several  factors obstruct a clearer insight into the phenomenon: 
there are many electronic bands crossing the Fermi level in most materials and
the AMR of a crystalline material has various contributions of different
symmetries. Ab initio calculations performed in FeNi\cite{Banhart:1995_a}
and FeCo\cite{Ebert:2000_a} disordered alloys agree reasonably well with
experimentally determined AMR but they do not allow for any detailed
conclusions about its mechanisms. On the other hand, the
model\cite{Mott:1964_a} of current-carrying $s$-states scattered to
spin-orbit-coupled $d$-states provides a relatively transparent picture of the
AMR\cite{Malozemoff:1986_a} but requires the fitting of one or more
phenomenological parameters and even then a clear-cut correspondence to ab
initio results for ferromagnetic transition metals could not be 
established.\cite{Banhart:1995_a}

\begin{figure}[b]
\begin{center}

\begin{tabular}{lll}
\\[-6mm]
  \includegraphics[scale=0.25]{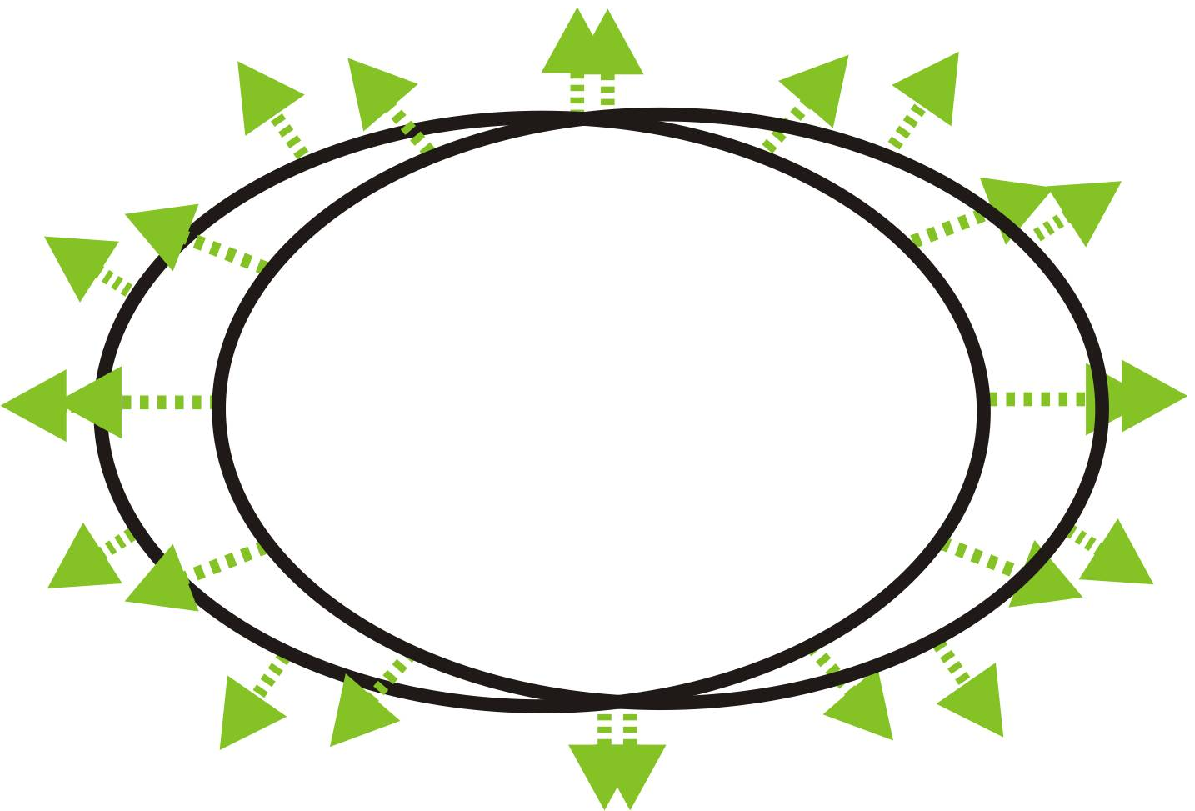} &
  \includegraphics[scale=0.25]{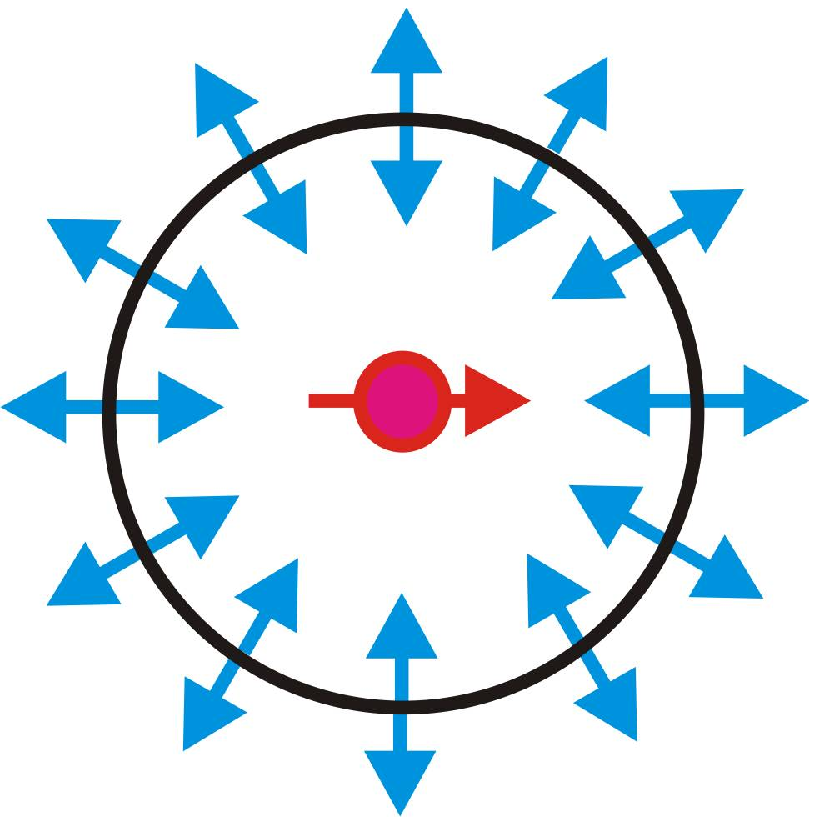} &
  \includegraphics[scale=0.25]{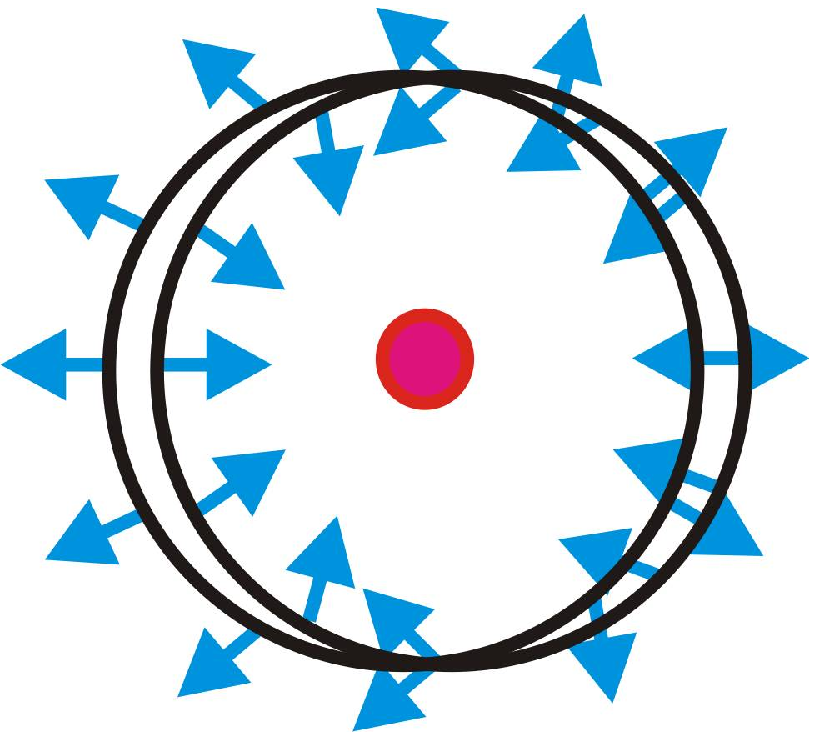} \\[-5mm]
(a)
 & (b) & (c) \\[-2mm]
\end{tabular}
\end{center}
\caption{The AMR can originate from three distinct mechanisms combining
  magnetization (pointing to the right in this sketch) and SOI, that break the
  isotropy: (a) anisotropic Fermi velocities (arrows) along the Fermi surface,
  or anisotropic relaxation rates due to
  (b) unpolarized bands (represented by the indicated 
  isotropic spin texture) scattered
  by anisotropic impurities or (c)   partially polarized bands scattered by
  isotropic impurities.} 
\label{fig-01}
\end{figure}

Diluted magnetic semiconductors, and (Ga,Mn)As in particular, offer a
promising system in which these issues become 
simplified:\cite{Jungwirth:2006_a} Fermi level lies close to the
top of the valence band so that only few bands are involved in transport, and
in addition, their SOI is strong.  Moreover, experiments done so far show
that the non-crystalline component 
of the AMR,\cite{Rushforth:2007_a,Limmer:2008_a} arising from the breaking of
the symmetry by choosing a specific current direction, outweighs the
crystalline components in most of the metallic highly Mn-doped materials.
In attempting to describe the AMR in such system, we can begin with a model
isotropic but still spin-orbit coupled band structure 
and add the effect of magnetization in the three possible distinct ways,
as sketched in Fig.~\ref{fig-01}. Either (a) the magnetization 
induces a magnetotransport anisotropy via the SOI already at the level of
group velocities of the exchange-split Fermi surfaces,\footnote{Thus induced
  anisotropy of Fermi surfaces can nicely be illustrated on Fig.~\ref{fig-01}.
  Consider the spin texture in Fig.~\ref{fig-01}(b) as a typical example of
  SOI effect. Exchange splitting due to magnetization oriented horizontally in
  Fig.~\ref{fig-01}, effectively acts as a Zeeman energy due to fictitious 
  'magnetic field' $\vek{B}$ with the same orientation, see
  Eq.~(\ref{eq-02}). The states with wavevector $\vek{k}||\vek{B}$ will be
  shifted upwards (downwards) in energy because their spin is parallel
  (antiparallel) to $\vek{B}$ while the states $\vek{k}\perp\vek{B}$ remain
  intact. This $\vek{k}$-anisotropic shift in energy can be translated into
  the splitting of Fermi wavevectors of the two originally degenerate bands.} 
(b) it may enter via
anisotropic scattering of the unpolarized spin-orbit coupled carriers from 
polarized magnetic impurities, or (c) via anisotropic scattering of partially
polarized carriers which does not require a magnetic character
of the scatterers. We point out that the mechanisms (a) and (c) represent a
situation where both fundamental ingrediences of the AMR (SOI and
magnetization) are present in the same states of the band structure. The SOI
is necessary for AMR to occur but, at the same time, it weakens the effect
of magnetization so that weaker AMR may be expected whenever the mechanisms
(a) or (c) dominate. On the other hand, in mechanism (b), the SOI in an
unpolarized carrier band can be strong while the magnetization of the
impurities remains at
100\%. Consequently, very large AMR can arise if this mechanism is
important.\cite{Trushin:2009_a}

We show in this paper that metallic
(Ga,Mn)As is a favourable system for the purposes of studying AMR. Not only because of its relatively simple
(effective) Hamiltonian (described in Sec.~\ref{sec-II}) and the dominance
of the AMR mechanism (b), but also because of the way the AMR model can be
simplified (as shown in Sec.~\ref{sec-III}) down to analytical formulae
revealing the basic AMR trends (see Sec.~\ref{sec-IV}).  This analysis is our
main result together with the detailed explanation of the AMR sign in
(Ga,Mn)As (resistance parallel to magnetization is smaller than perpendicular
to magnetization) which is observed
experimentally\cite{Rushforth:2007_a,Limmer:2006_a,Wang:2005_a,Goennenwein:2004_a,Matsukura:2004_a,Tang:2003_a,Jungwirth:2003_b,Baxter:2002_a}
and is opposite to most magnetic metals.\cite{Jaoul:1977_a,McGuire:1975_a} 
The results in Sec.~\ref{sec-IV} include analytically evaluated anisotropic
conductivity on several levels of model complexity, and the most simplified
model allows to clearly identify the physical mechanism that determines the
sign of the AMR in (Ga,Mn)As. Our
approach\cite{Jungwirth:2002_c} is based on the relaxation time
approximation (RTA). It is desirable to put the present results into more
precise terms by exactly solving the Boltzmann equation in its full integral
form as the authors did for the simpler Rashba system
recently.\cite{Vyborny:2008_a} However, we have also shown in detail that the
sign and the qualitative behaviour of AMR in model systems can be understood
using the RTA.\cite{Trushin:2009_a}  The explanation of the negative AMR
(resistance parallel to magnetization is smaller than perpendicular to
magnetization) provided here is robust against the choice of approximation.

\section{Basic model of AMR in metallic $\mbox{(Ga,Mn)As}$}
\label{sec-II}

Three principal ingredients, described in the following Subsections
(A,B,C), are necessary to model the conductivity and its magnetic
anisotropy: (A) The band structure yielding the spectrum and wave
functions, (B) the scattering mechanism, and (C) a transport formalism
which combines the former two and produces the conductivity
tensor. Given that we base our approach to (C) on relaxation-time
approximate solution to the semiclassical Boltzmann equation, we
basically need the Fermi velocities derived from the band dispersions,
and the relaxation times calculated from the spectrum, wave functions
and the relevant form of the impurity potential.

\subsection{Virtual-crystal kinetic-exchange model of (Ga,Mn)As bands}
\label{sec-IIA}

The valence-band kinetic-exchange model of (Ga,Mn)As with metallic
conductivities is an established qualitative and often
semiquantitative theoretical
approach.\cite{Jungwirth:2006_a,Jungwirth:2007_a} The description is
based on the canonical Schrieffer-Wolff transformation of the Anderson
Hamiltonian\cite{Schrieffer:1966_a} 
which for (Ga,Mn)As replaces hybridization of Mn $d$
orbitals with As and Ga $sp$ orbitals by an effective spin-spin
interaction of $(L=0;\, S=5/2)$ local moments with host valence-band
states. This step proves essential to effectively separate the
different AMR mechanisms (a,b,c), symbolized in Fig.~\ref{fig-01},
because --- except for the spin-spin interaction which will be treated
as we describe below --- it completely detaches the Mn states from the
spin-orbit coupled host-valence-band states.  These valence-band
states are conveniently parametrized by the Luttinger parameters
$\gamma_1,\gamma_2,\gamma_3$ and spin-orbit splitting $\Delta_{SO}$ in
the six-band Kohn-Luttinger
Hamiltonian\cite{Luttinger:1956_a,Baldereschi:1973_a} $H_{KL}$.  The local
interaction between Mn magnetic moments $\vek{S}_I$ (located at
$\vek{R}_I$) and valence hole spins $\vek{s}$ (at $\vek{r}$), being at
the root of the carrier-mediated ferromagnetism in (Ga,Mn)As, is the
kinetic exchange and it is described by single
parameter\cite{Okabayashi:1998_a,Jungwirth:2006_a} $J_{pd}$. In order
to model the band structure of (Ga,Mn)As including disorder electrical
potential $V$ associated with the Mn magnetic moments, we treat
the Hamiltonian 
\begin{eqnarray}\label{eq-01}
 H&=& H_{KL} + V_{dis} = \\[2mm] \nonumber
&& \hskip-0cm
      H_{KL}
      +J_{pd}\sum_{I} \vek{S}_I\cdot \vek{s}\, \delta(\vek{r}-\vek{R}_I)
      +\sum_{I} V(\vek{r}-\vek{R}_I)
\end{eqnarray}
by the virtual-crystal mean-field\cite{Abolfath:2001_a} approximation,
whence we get the single-particle Hamiltonian (in momentum representation) of
the (Ga,Mn)As valence band:
\begin{equation}\label{eq-02}
  H=H_{KL} + h \hat{e}_M\cdot \vek{s}\,.
\end{equation}
Here, $\hat{e}_M$ stands for the unit vector in the direction of the
mean-field magnetization, $h=J_{pd}N_{\mathrm{Mn}}S_{\mathrm{Mn}}$, and
the magnetic moment of Mn is $S_{\mathrm{Mn}}=5/2$.
In this paper, we will only consider substitutional Mn
with volume density $N_{\mathrm{Mn}}$ as in optimally annealed
samples,\cite{Jungwirth:2005_b} and assume zero temperature.
In the band structure model, we thus disregard the
randomness in the Mn distribution over the crystal and the ensuing spatial
inhomogeneity of the exchange interaction, and also we completely ignore the
disorder defined by the electrical potential $V$ in Eq.~(\ref{eq-01}) of every
single substitutional Mn which is an ionized acceptor.
Within this approximation, the effect of
the Mn atoms present in the crystal is reduced only to the effective
Zeeman-like term in~Eq.~(\ref{eq-02}) due to the kinetic exchange of the
valence holes with the Mn $d$-states.  Explicit form of the
$\vek{k}$-dependent $6\times 6$ matrix $H_{KL}$ in a convenient basis 
is given e.g. by Eq.~(A8) of Ref.~\onlinecite{Abolfath:2001_a}.

As we are aiming at a simple model of the non-crystalline AMR
component only, we will treat $H_{KL}$ in the spherical approximation,
implemented by setting $\gamma_2,\gamma_3$ to their average
value.\cite{Baldereschi:1973_a} In this approximation the dispersion
of all six valence bands becomes isotropic in the absence of the
kinetic-exchange field. The $6\times
6$ Hamiltonian~(\ref{eq-02}) can be diagonalized numerically and provide the
valence bands $E_{n}(\vek{k})$ of (Ga,Mn)As which are split by the exchange
field $h$. The index $n$ labels the two heavy hole bands ($n=1,2$),
two light hole bands
($n=3,4$), both of the $\Gamma_8$ symmetry and total angular momentum
$J=3/2$ in the $\Gamma$-point, and two split-off bands ($n=5,6$) with the
$\Gamma_7$ symmetry and $J=1/2$ in the $\Gamma$-point.  Note that spin is
not a good quantum number owing to the presence of SOI. Expectation value of
spin along any of the Fermi surfaces can be visualized as a spin texture
rather than having separate spin up and spin down bands. An example in
Fig.~\ref{fig-01}(b) that corresponds\cite{Rushforth:2007_b} to the
$n=1,2$ bands of Hamiltonian~(\ref{eq-02}) with $h\to 0$, shows that for each
$\vek{k}$ there are two states with opposite spin whose direction, however,
depends on $\vek{k}$, contrary to systems without SOI.

\subsection{Scattering on random Mn impurities}
\label{sec-IIB}

In order to get finite conductivity at zero temperature, we need to go beyond
the virtual-crystal concept of Eq.~(\ref{eq-02}). We follow
Ref.~\onlinecite{Ziman:1961_a} and use the Fermi golden rule (or first order
Born approximation treatment of $V_{dis}$) as the simplest model of scattering
to calculate the transport scattering rates $\Gamma_{n,\vek{k}}$ of the Bloch
states from Eq.~(\ref{eq-02}):
\begin{widetext}
\begin{eqnarray}\label{eq-03}
  \Gamma_{n,\vek{k}} &=& \frac{2\pi}{\hbar} N_{\mathrm{Mn}}
    \times \sum_{n'} \int \frac{d^3k'}{(2\pi)^3}
  |M^{\vek{k}\vek{k}'}_{nn'}|^2 
  \delta\big(E_n(\vek{k})-E_{n'}(\vek{k}')\big) 
  (1-\cos \theta_{vv'})\,,
\end{eqnarray}
\end{widetext}
where we use\cite{Ziman:1961_a} $\theta_{vv'}$, 
the angle subtended by the velocities
$\vek{v}_n(\vek{k})$ and $\vek{v}_{n'}(\vek{k}')$
to take into account $\vek{v}$ which need
not be parallel to $\vek{k}$ in case the combined effect of the SOI and
magnetization distorts the Fermi surfaces as suggested by the sketch 
in Fig.~\ref{fig-01}(a).

Substitutional Mn act as acceptors and their magnetic moments participate in
the ferromagnetic order of (Ga,Mn)As.
Acknowledging the magnetic and non-magnetic part of $V_{dis}$, we take
\begin{equation}\label{eq-04}
M^{\vek{k}\vek{k}'}_{nn'} =\langle z_{\vek{k}'n'}|M^B + M^C|z_{\vek{k}n}\rangle
\end{equation}
for the scattering matrix elements between two eigenstates of the 
Hamiltonian~(\ref{eq-02}). In the six-band notation of Eq.~(\ref{eq-02}), 
the magnetic part of a single Mn impurity scattering
operator is
\begin{equation}\label{eq-04b}
  M^B = J_{pd}S_{\mathrm{Mn}} \hat{e}_M\cdot \vek{s}\,,
\end{equation}
corresponding to the second term in Eq.~(\ref{eq-01}). Explicit form of the
spin $6\times 6$ matrices $\vek{s}$ is again given in
Ref.~\onlinecite{Abolfath:2001_a}.  The non-magnetic part
$M^C$ describes screened Coulomb attraction of the valence holes to the
ionized acceptors and we therefore take
\begin{equation}\label{eq-04c}
  M^C = V(|\vek{k}-\vek{k}'|)\,\mathds{1}\,, \quad
  V(q)=-\frac{e^2}{\varepsilon} \frac{1}{q^2+q_{TF}^2}\,,
\end{equation}
where $\mathds{1}$ denotes a $6\times 6$ unity matrix,
$\varepsilon$ is the host
semiconductor dielectric constant, $q_{TF}=\sqrt{e^2 g/\varepsilon}$ the
Thomas-Fermi screening wavevector,\cite{Ashcroft:1976_a,Jungwirth:2002_c} and
$g$ the density of states at the Fermi level.

It is important that the two scattering operators~(\ref{eq-04b},\ref{eq-04c})
add up 'coherently' in Eq.~(\ref{eq-04}). The 'incoherent' sum
$|M^{\vek{k}\vek{k}'}_{nn'}|^2 =
 |\langle z_{\vek{k}'n'}|M^B|z_{\vek{k}n}\rangle|^2
+|\langle z_{\vek{k}'n'}|M^C|z_{\vek{k}n}\rangle|^2$
would describe a physically different situation with
two distinct types of scatterers, magnetic and non-magnetic ones. Such
incoherent sum, with appropriately defined scattering operators, was used
earlier\cite{Jungwirth:2002_c} to describe more realistic (Ga,Mn)As systems
that contain interstitial Mn atoms or As antisites in addition to the
substitutional Mn. 

To summarize our model description of substitutional Mn impurities in
GaAs, the Mn atoms in Ga$_{1-x}$Mn$_x$As enter our model at three
different places: (i) As acceptors and in the absence of other dopants
they determine the Fermi level $E_F$ and therefore the density of
states.  (ii) On the virtual crystal approximation (VCA) level, they
cause the ferromagnetic-exchange splitting of the hole bands, and
(iii) because of the random distribution in the lattice, the Mn
impurities also cause scattering. The essential feature of the Mn
impurity potentials for the AMR is that they contain components which
are proportional to the Mn local moments and that these moments are
ordered in the ferromagnetic state, as expressed in Eq.~(\ref{eq-04b}).
We stress that considering (ii) and (iii) simultaneously leads to only
a small "double-counting" error in the description of the effect of
the Mn-related impurity potential.
In terms of the VCA, we assume in (ii) that each site on the cation
(Ga) sublattice of the host semiconductor is occupied by a mixture of
$x$ Mn and $1-x$ Ga. This yields an effective mean potential which
shares the full periodicity of the host zinc-blende lattice. Strictly
speaking, the scattering potential of randomly distributed Mn on the
cation sublattice should be described as the difference between the
full impurity potential due to Mn and the above VCA
potential. Similarly the remaining sites occupied by Ga should be
described by the difference between the Ga potential and the VCA
potential. Ignoring the latter difference and taking the full Mn
impurity potential for sites occupied by Mn when describing scattering
in (iii) is therefore not a precise procedure but it introduces only a
small error for Mn dopings not exceeding several per cent.

\subsection{Conductivity of (Ga,Mn)As in the relaxation time approximation}

We now calculate the conductivity tensor using the semiclassical formula based
on the Boltzmann transport equation:\cite{Jungwirth:2002_c}
\begin{equation}\label{eq-05}
  \sigma_{ij} = e^2 \sum_n \int \frac{d^3 k}{(2\pi)^3}
  (\hbar \Gamma_{n\vek{k}})^{-1} v^i_n(\vek{k}) v^j_n(\vek{k}) 
  \delta(E_F - E(\vek{k}))\,.
\end{equation}
We assume zero temperature hence the conductivity is determined exclusively by
states on the Fermi level $E_F$. The Fermi velocities 
are calculated as $\vek{v}_n(\vek{k}) = (1/\hbar)\nabla_{\vek{k}} E_n$. 
The conductivity tensor depends on the direction of magnetisation
$\hat{e}_M$ through Eqs.~(\ref{eq-02})~and~(\ref{eq-04b}), that is 
owing to the combined effect of magnetisation and SOI. 
Generally, the tensor~(\ref{eq-05}) may be non-diagonal and the
resistivity tensor $\rho_{ij}$ is its inverse.

Here we consider a 'pure-AMR-configuration' where $\sigma_{ij}$ is
symmetric, i.e. free of any antisymmetric Hall components. It can be
envisaged as a Hall-bar device fabricated from a thin (Ga,Mn)As film
with an arbitrary in-plane magnetization. Experimentally, the
magnetization will be controlled by a weak magnetic field whose direct
effect on the AMR will be neglected. The longitudinal and
transverse voltage drops are proportional to $\rho_{xx}$ and
$\rho_{xy}$; spherical approximation (see Sec.~\ref{sec-IIA}) makes
both the orientation of the film and of the Hall bar device with
respect to the crystallographic axes irrelevant. 
In an out-of-plane configuration, 
the resistivity acquires an antisymmetric anomalous Hall component, see
Ref.~\onlinecite{Sinitsyn:2007_a} for a comprehensive review.  For the
in-plane configuration considered here, the resistivity is completely symmetric,
$\rho_{xy}(-\hat{e}_M)=\rho_{xy}(\hat{e}_M)=\rho_{yx}(\hat{e}_M)$, and
obeys $\rho_{xy}(\phi)/\rho_{av}=C_I\sin 2\phi$ where $\phi$ is the
angle between magnetization and current (Hall bar device) direction.
This is a result of a general analysis\cite{Ranieri:2008_a} of isotropic
systems with symmetry broken by the current flow. According to this analysis,
the diagonal resistivity, $\rho_{xx}(\phi)=\rho_{av}(1+C_I\cos 2\phi)$,
carries the same information about AMR as $\rho_{xy}(\phi)$, which is
concentrated into the non-crystalline AMR coefficient $C_I$ 
[$\rho_{av}$ is the angular average of $\rho_{xx}(\phi)$].

We use the following definition of the AMR:
\begin{equation}\label{eq-07} 
  \mathrm{AMR} \equiv
  -2\frac{\sigma_{||}-\sigma_{\perp}}{\sigma_{||}+\sigma_\perp}
       = 2\frac{\rho_{||}-\rho_{\perp}}{\rho_{||}+\rho_\perp}\,,
\end{equation}
where $\sigma_{\parallel}$ and $\sigma_\perp$ ($\rho_{\parallel}$ and
$\rho_\perp$) are the longitudinal conductivities (resistivities) for
current parallel and
perpendicular to the magnetization, respectively. Note that the
off-diagonal resistivities vanish for $\phi=0$ or $90^\circ$.
For practical purposes, we
can set $\hat{e}_M$ in the $x$-direction and $\sigma_{\parallel}\equiv
\sigma_{xx}$, $\sigma_{\perp}\equiv \sigma_{yy}$ as long as we stay with the
spherical approximation. The AMR of
Eq.~(\ref{eq-07}) equals to $2C_I$.

\begin{figure*}
\begin{tabular}{cc}
  \kern-1cm\hbox{\includegraphics[scale=1.]{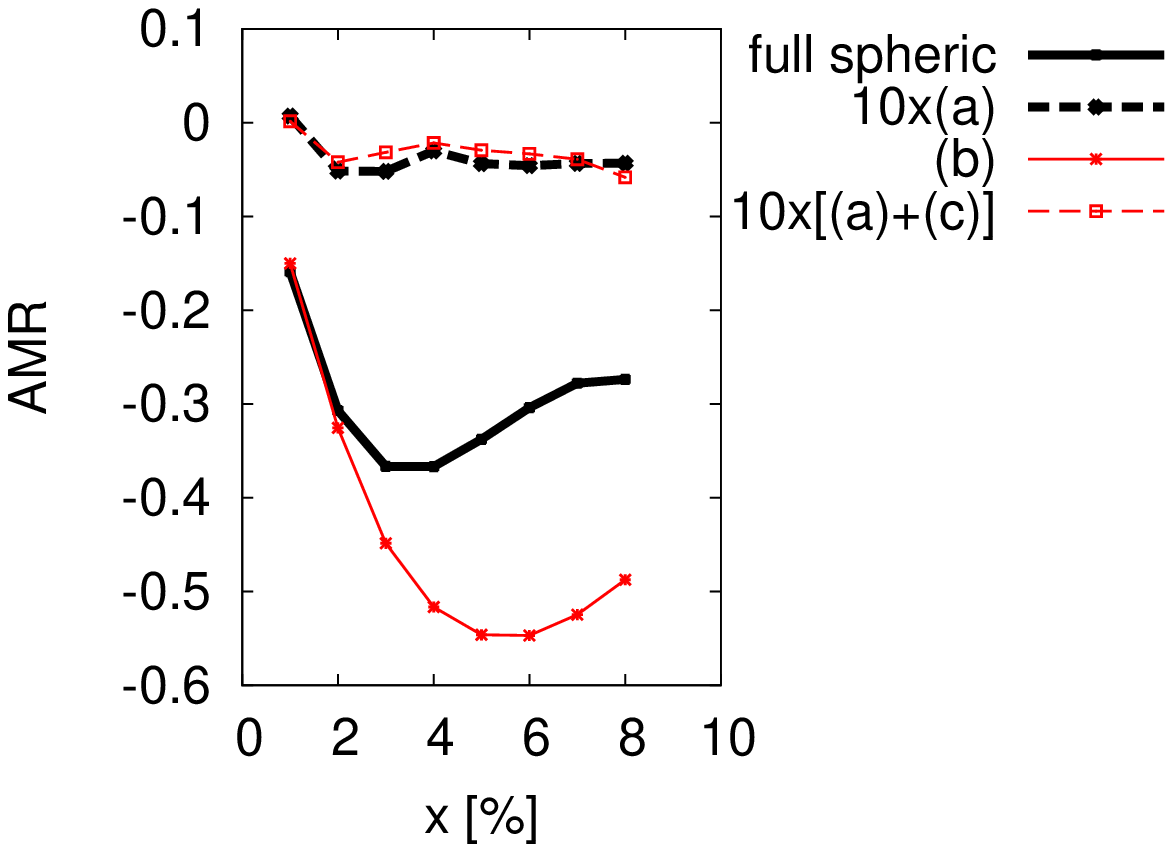}} &
  \hskip-3.1cm\includegraphics[scale=1.]{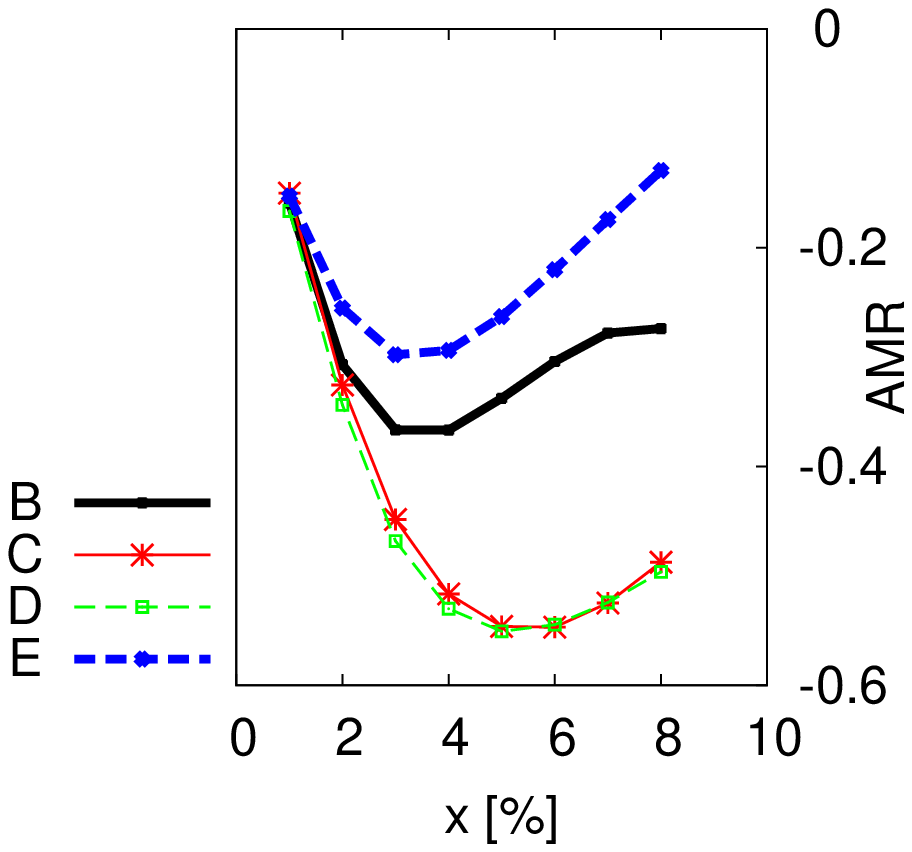} 
\end{tabular}
\caption{{\em Left:} 
study of the influence of the three mechanisms (a,b,c) sketched in
  Fig.~\ref{fig-01} on the total AMR. The model in spherical approximation as
  of Sec.~\ref{sec-II}, serves as a reference ('full spheric'). Note that the
  upper two curves are upscaled by a factor of ten. {\em Right:}
  subsequent approximations (see text) aiming towards an analytically
  solvable model. Note that (B) and (C) is the same as 'full spheric' and (b)
  on the left panel, respectively, while (D) and (E) correspond to gradual
  decoupling of the light holes.}
\label{fig-04}
\end{figure*}

\section{Non-crystalline AMR of heavy holes in the spherical approximation}
\label{sec-III}

In this Section, we first show that within our description of metallic
(Ga,Mn)As samples, the AMR trends are governed by the mechanism sketched in
Fig.~\ref{fig-01}(b). Then we proceed to showing that only heavy hole bands
need to be considered when analyzing the basic AMR characteristics in
(Ga,Mn)As on a qualitative level.

\subsection{Origins of anisotropy}

Let us consider how the conductivity in Eq.~(\ref{eq-05}) can become
magnetization-dependent.  In agreement with the intuitive analysis of
Fig.~\ref{fig-01}, magnetization direction $\hat{e}_M$ can enter
Eq.~(\ref{eq-05}) either via (a) the group velocity components $v^i_n$ or
(b,c) the scattering rates $\Gamma_{n\vek{k}}$.  Considering
Eqs.~(\ref{eq-03},\ref{eq-04}), the scattering rates may depend on $\hat{e}_M$
either through (b) the scattering operator ($M_B+M_C$ in our specific case) or
(c) the wavefunctions $|z_{\vek{k}n}\rangle$ and/or energies $E_n(\vek{k})$
and density of states of the carrier bands. The last mechanism, for example,
lies at the heart of the $s$-$d$ model of AMR in transition
metals\cite{Malozemoff:1986_a} where isotropic and spin-independent scattering
operators have been assumed.\cite{McGuire:1975_a,Jaoul:1977_a} The anisotropy
arises due to the competition of SOI and magnetization which splits the five
$d$ states according to $l_M$, their angular momentum projection along
$\hat{e}_M$. These states play the role of $|z_{\vek{k}'n'}\rangle$ in
Eq.~(\ref{eq-04}), and because of their $l_M$-dependent spatial form they
cause $\hat{e}_M$-dependent scattering rates $\Gamma_{n\vek{k}}$ in the
$s$-states that carry the current. Plugged back into Eq.~(\ref{eq-03}), these
anisotropic scattering rates may lead to different $\sigma_{xx}$ for
$\hat{e}_M$ parallel and perpendicular to the $x$-direction, i.e.
$\sigma_{\parallel}\not=\sigma_\perp$. 

In (Ga,Mn)As, we are going to take advantage of the tunability that the
effective model outlined in Sec.~\ref{sec-II} offers: we will switch the
particular mechanisms (a,b,c) on and off to see how important they are for the
total AMR. We use the band-structure model as described in Sec.~\ref{sec-IIA}
(termed 'full spherical') as a reference.  Within this full spherical model,
the calculated AMR as a function of Mn doping $x$ is negative in the
considered range between 2 and 10\% and its magnitude reaches a clear maximum
as shown by the middle curve in the left panel of Fig.~\ref{fig-04} (the
maximum is related to the competition between the electric and magnetic parts
of the scattering operator as we explain in Sec.~\ref{sec-IV}).  To see the
effect of the mechanism (b) alone, we set $h=0$ in Eq.~(\ref{eq-02}) but leave
the scatterer anisotropy unchanged by keeping nonzero $J_{pd}$ in
Eq.~(\ref{eq-04b}).  The bottom curve in the left panel of Fig.~\ref{fig-04}
demonstrates that the AMR quantitatively changes within a factor of two but
its overall form remains the same.

On the other hand, the result alters dramatically when we switch off the
anisotropy in the scattering operator [mechanism (b)] or the anisotropy in
relaxation rates as a whole [mechanisms (b) and (c) together].  The former is
accomplished by setting $M_B=0$ in Eq.~(\ref{eq-04}), the latter is done by
replacing $\Gamma_{n,\vek{k}}$ by a constant whose value is irrelevant because
it cancels out in Eq.~(\ref{eq-07}). In both cases, we obtain AMR that is more
than an order of magnitude smaller than for the full spherical model, see
Fig.~\ref{fig-04}. This result indicates that mechanism (b) is a crucial part
of the AMR model of metallic (Ga,Mn)As, the other mechanisms (a,c) provide
only quantitative corrections and when kept alone without mechanism (b), they
produce negligible AMR.

\subsection{Heavy holes}

The next simplification of the model we make in order to provide a simple
physical picture of the AMR in (Ga,Mn)As, is to neglect the light holes. We
can accomplish this in two steps: we first discard the current carried by the
light holes, i.e. sum in Eq.~(\ref{eq-05}) over $n=1,2$ only, and then also
disable the scattering from the heavy hole to light hole bands, i.e. sum in
Eq.~(\ref{eq-03}) over $n'=1,2$ only. Numerical calculation again shows that
this procedure does not alter the qualitative behaviour of the AMR. All levels
of approximations are summarized in the right panel of Fig.~\ref{fig-04}. For
completeness, we start with the 'full spheric' reference (curve labelled by
'B') as on the left panel and proceed to suppressing the band polarization by
putting $h=0$ in Eq.~(\ref{eq-02}) which yields the curve 'C'.  Omission of
light-holes from current-carrying states in the transport equation 
but not from the final states in the scattering matrix elements
produces the data labelled by 'D', and the completely heavy-hole-only model
(six-band model where the other four bands are disregarded) is denoted by 'E'.

For further studies of this model it may be interesting that the difference
between curves 'C' and 'D' is remarkably small, in other words, the anisotropy
of the light hole transport is almost identical to that of the heavy holes,
provided we have set $h=0$ in Eq.~(\ref{eq-02}). However, the main conclusion
of this Section is that, within the studied range of Mn doping, the AMR is
determined by the anisotropy of the relaxation rates of heavy holes induced by
the Mn electro-magnetic scatterers. This allows us to derive an approximate
analytical formula for the AMR in (Ga,Mn)As which we discuss in the following
section.

\section{Qualitative analytical results for AMR in $\mbox{(Ga,Mn)As}$}
\label{sec-IV}

We first provide analytical expressions for the AMR corresponding to the 
curve~'E' in Fig.~\ref{fig-04} with an additional approximation
that the current $\vek{I}$ is proportional only to the transport
life-time of carriers with $\vek{v}_n \parallel \vek{I}$.

Using the explicit form of the heavy hole wavefunctions given in
Ref.~\onlinecite{Rushforth:2007_b}, the transport scattering rates of
Eq.~(\ref{eq-03}) corresponding to $M^B+M^C$ of
Eqs.~(\ref{eq-04b},\ref{eq-04c}) can be evaluated as
\begin{widetext}
\begin{eqnarray}\label{eq-15a}
  \Gamma_{\pm}(0) &=& \frac{\pi}{2}  \left[ \frac{a^2}{k_F^2}
  \bigg(12+18b+\frac{8}{b-1} -
  (1+3b)^2\ln\frac{b+1}{b-1}\bigg) \mp 
a\bigg(4+6b+6b^2- (b+1)(1+3b^2)\ln\frac{b+1}{b-1}\bigg)
   +2 k_F^2 \right]\,,
\end{eqnarray}
where $\Gamma_{\pm}(\phi)$ are the scattering rates of the two heavy-hole
bands with
$\phi$ equal to the angle between $\vek{k}$ and $\hat{e}_M$. 
Next, we get 
\begin{eqnarray}\label{eq-15b}
  \Gamma_{\pm}(\frac{1}{2}\pi) &=& 
\frac{\pi}{2} \left[
  \frac{1}{3}k_F^2 + \frac{a^2}{k_F^2} \bigg( 12+18b + \frac{8}{b-1}
  -(1+3b)^2\ln\frac{b+1}{b-1}\bigg)\right]\,.
\end{eqnarray}
\end{widetext}
We used shorthands $a=-(e^2/2\varepsilon k_F^2)/J_{pd}S_{\mathrm{Mn}}$ with
$k_F$ denoting the common Fermi wavevector of the heavy holes, and
$b=-(1+q_{TF}^2/k_F^2)$.

Conductivities are evaluated as 
\begin{eqnarray}\label{eq-13a}
\sigma_{\parallel} &\propto& \displaystyle
                      \frac{1}{\Gamma_+(0)}  +\frac{1}{\Gamma_-(0)}\\  
\label{eq-13b}
\sigma_{\perp} &\propto& \displaystyle 
\frac{1}{\Gamma_+(\textstyle\frac{1}{2}\pi)}+
\frac{1}{\Gamma_-(\textstyle\frac{1}{2}\pi)}=
                      \frac{2}{\Gamma_+(\textstyle\frac{1}{2}\pi)}\,,
\end{eqnarray}
and yield AMR which is up to an overall factor of $\approx 2$ the same as 
if we performed the complete $\vek{k}$-integration in Eq.~(\ref{eq-05}), see 
Fig.~\ref{fig-05}(a). The approximation of taking into account only states with
$\vek{k}\parallel\vek{I}$ is therefore qualitatively valid.

\subsection*{Short range scatterers}

Results of Eqs.~(\ref{eq-15a},\ref{eq-15b}) are analytical but the formula for
AMR is rather complicated. We can attain a clear insight into how the observed
AMR trends arise if we further simplify the model. We replace $M^B+M^C$
of Eqs.~(\ref{eq-04b},\ref{eq-04c}) by another scattering operator
\begin{equation}\label{eq-08}
  M^{s.r.} \propto \hat{e}_M\cdot \vek{s} + \alpha \mathds{1}\,.
\end{equation}
That is, we assume $\vek{k}$-independent electric part of the scattering
potential.

\begin{figure}
\begin{tabular}{ll}
  \hskip-1cm\includegraphics[scale=0.54]{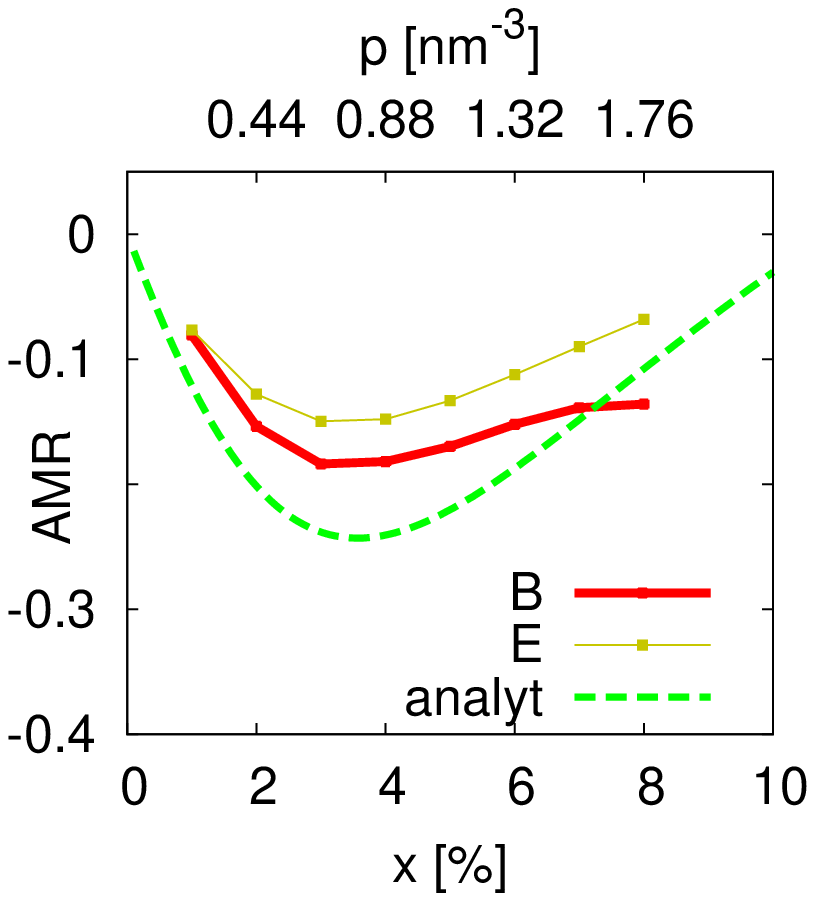}  &
  \hskip-.7cm\includegraphics[scale=0.54]{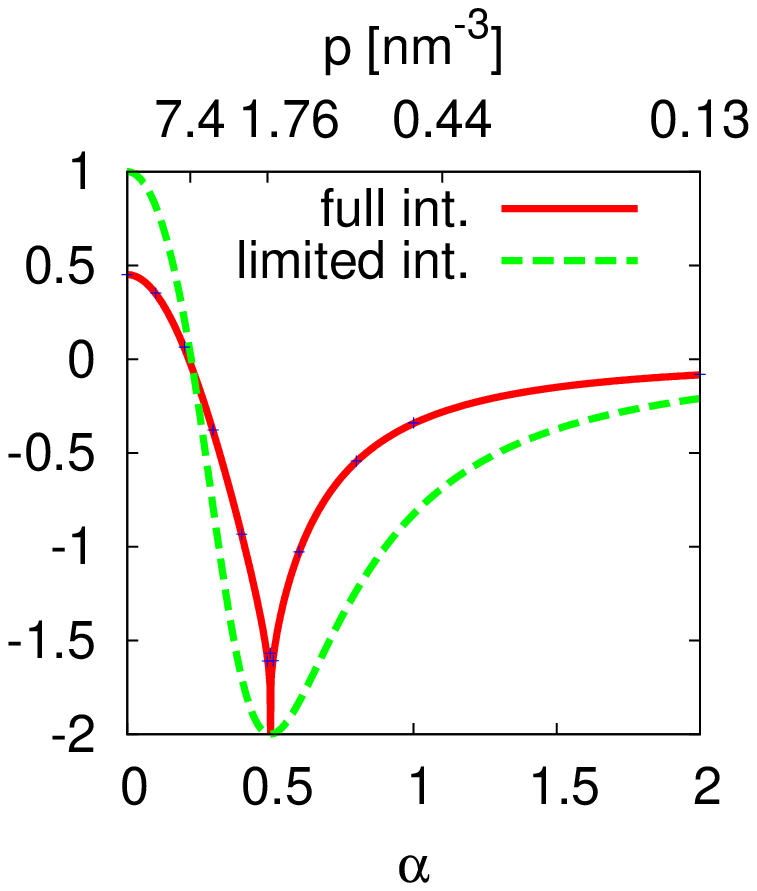}  \\[-2mm]
  (a) & (b)
\end{tabular}
\caption{  (a) The analytical model of AMR based on 
  Eqs.~(\ref{eq-15a},\ref{eq-15b}). Curves 'B', 'E' are taken from
  the right panel of 
  Fig.~\ref{fig-04}. Estimates of the hole concentrations shown on the
  upper axis assume one hole per Mn and neglect the light-hole bands.
  (b) The AMR as a function of the effective Coulomb scattering
  strength given by Eq.~(\ref{eq-11}). The 'limited-integration' 
  model (only states with $\vek{k}$ parallel to
  the current) is represented by Eq.~(\ref{eq-14}), the integrated model where
  all states contribute to the current is described after Eq.~(\ref{eq-10}).
}
\label{fig-05}
\end{figure}

The transport scattering rates of Eq.~(\ref{eq-03}) again depend only on the
angle $\phi$,
\begin{equation}\label{eq-09}
\Gamma_{\pm}(\phi)
  \propto \frac{1}{6}\cos^2\phi \pm \alpha\cos\phi +
                        \alpha^2 + \frac{1}{12}\,,
\end{equation}
and specifically for states with $\vek{k}$ parallel and perpendicular to
$\hat{e}_M$, we get
%
\begin{equation}\label{eq-12}
\Gamma_{\pm}(0) \propto (\alpha\pm\frac{1}{2})^2\,, \qquad
\Gamma_{\pm}(\frac{1}{2}\pi)\propto \alpha^2+\frac{1}{12}\,.
\end{equation}
Conductivities evaluated using Eqs.~(\ref{eq-13a},\ref{eq-13b}) 
simplify to
%
\begin{eqnarray}\label{eq-16a}
\sigma_{\parallel} &\propto& 
          \frac{1}{(\alpha+\frac{1}{2})^2} + \frac{1}{(\alpha-\frac{1}{2})^2}\\
\label{eq-16b}
\sigma_{\perp} &\propto& \displaystyle
                      \frac{2}{\Gamma_+(\textstyle\frac{1}{2}\pi)} =
                     \frac{2}{\alpha^2+\frac{1}{12}}\,,
\end{eqnarray}
which gives, using Eq.~(\ref{eq-07}), our previous 
result\cite{Rushforth:2007_a,Rushforth:2007_b}
\begin{equation}\label{eq-14}
  \mathrm{AMR} = -\frac{20\alpha^2-1}{24\alpha^4 - 2\alpha^2 + 1}\,.
\end{equation}
In order to link this result, plotted in Fig.~\ref{fig-05}(b), to the previous
one that is based on the full form of the Coulomb scattering operator given by
Eq.~(\ref{eq-04c}), we need to focus on the parameter $\alpha$. It represents
the effective strength of the electric part relative to the magnetic part of
the scattering potential of the Mn ions.
We can estimate $\alpha$ as an average over the Fermi surface
of the more realistic $V(|\vek{k}-\vek{k}'|)$ from Eq.~(\ref{eq-04c}),
\begin{equation}\label{eq-11}
  \alpha = \frac{\langle V\rangle_{FS}}{J_{pd}S_{\mathrm{Mn}}}=
  \frac{e^2 /\varepsilon}{J_{pd} S_{\mathrm{Mn}}}\cdot
  \frac{1}{4k_F^2} \ln \left(1+\frac{4k_F^2}{q_{TF}^2}\right)\,.
\end{equation}
Explicitly, $\langle V\rangle_{FS}\equiv (4\pi
k_F^2)^{-1}\int_{FS} d^2 k' V(|\vek{k}-\vek{k}'|)$ with $\vek{k}$ fixed to an
arbitrary Fermi wave vector, $|\vek{k}|=k_F$.  The integral is taken over the
Fermi surface.

We thus find that the relative strength of the Coulomb scattering increases
with decreasing hole density, $p\approx k_F^3 /(3\pi^2)$, due to the screened
long-range nature of the Coulomb interaction which contrasts the short range of
the magnetic scattering. Namely, the momentum transfers $q$ scale down with
$k_F$ and we increasingly get to feel the singularity of $V(q)$ at $q=0$ even
though it is rounded by the screening, see Eq.~(\ref{eq-04c}).  Assuming a
value $p=0.5\unit{nm^{-3}}$ which is realistic in (Ga,Mn)As, we have
$k_F\approx 2.5\unit{nm}^{-1}$, $q_{TF}\approx 1.6\unit{nm}^{-1}$, and
$\alpha\approx 1.0$.

Equation~(\ref{eq-11}) provides the link between $\alpha$ and $p$ which is
determined by $x$, i.e. between horizontal axes of Figs.~\ref{fig-05}(a)
and (b). Qualitatively, increasing $x$ and thus also increasing $k_F$ implies
decreasing $\alpha$ because the scattering rate due to the magnetic part
scales $\propto k_F^2$ while the scattering rate due to the Coulomb scattering
given by $M^C$ of Eq.~(\ref{eq-04c}) grows considerably slower. If we estimate
$V(|\vek{k}-\vek{k}'|)$ by $V(2k_F)$, the growth will be only $\propto
k_F^2/(k_F^2a^2+k_Fa/\pi)$ where $a$ is the effective Bohr radius in
GaAs. Behaviour of the more sophisticated estimate used in Eq.~(\ref{eq-11}) is
qualitatively the same.

The simple form of scattering rates in Eq.~(\ref{eq-09}) allows to 
analytically perform the full integration over $\vek{k}$ in Eq.~(\ref{eq-05})
and to obtain more precise results for conductivity.
For $\sigma_{\parallel} =\sigma_{\parallel}^+ + \sigma_{\parallel}^-$  we get
\begin{equation}
\label{eq-10}
\sigma_{\parallel}^\pm \propto \int_{FS} 
\frac{d^2k \cos^2\phi}{\Gamma_{\pm}(\phi)}\,.
\end{equation}
The proportionality factor is
$e^2$ times squared Fermi velocity times the density of states at the Fermi
level.  Conductivities of the two bands are equal, so that
\begin{eqnarray}\label{eq-17}
\sigma_{\parallel} &\propto& 12 - 
36\alpha\ln\left|\frac{\alpha + \frac{1}{2}}{\alpha - \frac{1}{2}}\right|
+\\ \nonumber
 && \hskip-1cm \sqrt{18}\frac{24\alpha^2-1}{\sqrt{|6\alpha^2-1|}} 
\arcsinh \frac{\sqrt{|6\alpha^2-1|/18}}{|\alpha^2-\frac{1}{4}|}\,,\quad
\alpha^2>\frac{1}{6}\,,
\end{eqnarray}
and the same result with $\arcsinh$ replaced by $\arcsin$ applies 
for $\alpha^2<\frac{1}{6}$.

In order to evaluate $\sigma_{\perp}$, we employ a straightforward identity 
$\sigma_{\parallel}^\pm + 2\sigma_{\perp}^\pm = T^{\pm}$ with
$$
  T^{\pm} = \frac{2\sqrt{18}}{\sqrt{|6\alpha^2-1|}} 
\arcsinh \frac{\sqrt{|6\alpha^2-1|/18}}{|\alpha^2-\frac{1}{4}|}\,,\qquad
\alpha^2>\frac{1}{6}\,,
$$
and again with $\arcsin$ for $\alpha^2<\frac{1}{6}$, which can be derived by
inserting $\cos^2\phi=1-\sin^2\phi$ into Eq.~(\ref{eq-10}). The AMR evaluated
using Eq.~(\ref{eq-07}) is shown as the solid curve in Fig.~\ref{fig-05}(b).

Both models with simplified scattering operator $M^{s.r.}$ exhibit
qualitatively the same behaviour of AMR as a function of $\alpha$: positive
AMR for $\alpha$ close to zero, sign change to negative AMR 
already for a small value of $\alpha$, maximum AMR magnitude
at $\alpha=1/2$, and vanishing AMR for $\alpha\to\infty$. A comparison between
Figs.~\ref{fig-05}(a) and (b) reveals that the outstanding feature of the AMR
at $|\alpha|=1/2$ makes its way\footnote{Note that once we have set $h=0$ in
  Eq.~(\ref{eq-02}), the theoretical AMR 
  depends on the Mn doping $x$ only through the carrier density
  $p[\mathrm{nm}^{-3}]=0.22\times x[\%]$,
  because $N_{\mathrm{Mn}}$ appearing in Eq.~(\ref{eq-03}) drops out in
  Eq.~(\ref{eq-07}). This fact should not be taken {\em ad absurdum}
  ($N_{\mathrm{Mn}}\to 0$) because we still assume that 
  the substitutional Mn impurities provide the dominant source of scattering.
  Presence of other concurrent types of impurities (which is important when
  we compare the calculated AMR to experiments\cite{Jungwirth:2002_c}) 
  will make the AMR depend on
  the ratio between their concentration and $N_{\mathrm{Mn}}$. Nevertheless,
  this interplay will not simply obey Matthiessen's rule since we deal with
  anisotropic systems.\cite{Dugdale:1967_a} }
up to the full spherical model where it is
becomes broadened to the wide maximum (around $x\approx 4\%$ in
Fig.~\ref{fig-04}). Based on the model that employs $M^{s.r.}$ of
Eq.~(\ref{eq-08}), we now analyze the origin of the very large negative AMR at
$|\alpha|=1/2$. 

This maximum AMR value follows from the diverging conductivity
$\sigma_{\parallel}$ given by Eqs.~(\ref{eq-16a}) or~(\ref{eq-17}) which is
caused by vanishing scattering rate $\Gamma_\mp(0)$ for $\alpha=\pm 1/2$. The
scattering matrix element in Eq.~(\ref{eq-04}) vanishes for any $n',\vek{k}'$,
and so does the integral~(\ref{eq-05}), when $|z_{\vek{k}n}\rangle$ is an
eigenstate to $M^{s.r.}$ with eigenvalue zero; for illustrative purposes,
consider $M^{s.r.}=s_x+\alpha$, i.e. $\hat{e}_M$ pointing to the right in terms
of Fig.~\ref{fig-01}. The absence of scattering occurs for $\alpha=1/2$ when
$\vek{k}$ is parallel to the $x$-direction, as evident\cite{Rushforth:2007_b}
from the spin texture in Fig.~\ref{fig-01}(b).  Such infinite conductivity
will of course not occur in realistic systems; as soon as the dependences on
momentum transfer of the electric and magnetic parts of the scattering
potential will not be exactly the same, the eigenstate property of
$|z_{\vek{k}n}\rangle$ is lost. This is the case when we replace $M^{s.r.}$ by
the original $M^B+M^C$ of Eqs.~(\ref{eq-04b},\ref{eq-04c}).  However, as
Fig.~\ref{fig-05}(a) shows, there still remains a maximum in
$\sigma_{\parallel}$ which translates into a maximum of $|\mbox{AMR}|$ as a
fingerprint of the original $\sigma_{\parallel}$ singularity.

Positive AMR for purely magnetic scattering ($\alpha=0$) that follows from
Eq.~(\ref{eq-14}) can also be understood using the basic properties of
spin-3/2 states.\cite{Trushin:2009_a} The idea is that $|z_{\vek{k}n}\rangle$
is an eigenstate (with non-zero eigenvalue) to $\hat{e}_M\cdot \vek{s}$ for
$\vek{k}$ oriented along $\hat{e}_M$ thereby allowing scattering to a state
with $-\vek{k}$ that contributes strongly to the transport relaxation rate of
Eq.~(\ref{eq-03}). On the other hand, matrix elements of $\hat{e}_M\cdot
\vek{s}$ vanish\cite{Rushforth:2007_b,Trushin:2009_a} between states with
$\vek{k},\vek{k}'\perp\hat{e}_M$ and scatterring and resistivity are therefore
suppressed for the magnetization perpendicular to the current.  Relation
$\sigma_{\perp}>\sigma_{\parallel}$ ($\mbox{AMR}>0$) for $\alpha=0$, obtained
from the above analysis of the leading scattering channels which contribute to
transport, is confirmed by calculations that even take into account all final
states $\vek{k}'$ for the scattering and include the integration over all
$\vek{k}$-states in Eq.~(\ref{eq-05}).  The sign of the AMR changes quickly
from positive to negative when adding the electric component of the scattering
potential. In the model of short range potentials, summarized by
Eq.~(\ref{eq-14}), it occurs at $\alpha=1/\sqrt{20}$. Recall that for
realistic impurities with screened long-range Coulomb potential the transition
can be triggered by changing hole density, as explained in the discussion of
Fig.~\ref{fig-04}.

At the end of this Section, we recall that all considerations in this work are
based on the RTA. It has been shown in Ref.~\onlinecite{Vyborny:2008_a} that
the RTA does not provide the exact solution to the Boltzmann equation. Even in
the simple case of a spherical model in (Ga,Mn)As we should solve an integral
equation rather than merely evaluate integrals as in
Eq.~(\ref{eq-03}). Nevertheless, it has also been argued in
Ref.~\onlinecite{Trushin:2009_a} that the basic qualitative trends of the AMR
can be found already on the level of the RTA. In particular this applies to
the maximum of $|\mbox{AMR}|$ at $|\alpha|=1/2$ whose robustness was verified
by comparing the RTA and the exact solution of the Boltzmann equation in its
integral form on model Rashba-Dresselhaus SOI systems.

\section{Summary}

Non-crystalline anisotropic magnetoresistance (AMR) is governed only by the
angle between magnetization and current rather than by their orientation with
respect to crystallographic axes, as opposed to the crystalline AMR.  The
physical origin of the AMR is the combination of spin-orbit interaction (SOI)
and of the broken symmetry due to the presence of magnetization.  More
specifically, three distinct mechanisms may lead to the non-crystalline AMR:
Fermi surfaces distorted from the spherical shape that imply anisotropic Fermi
velocities, and anisotropic scattering rates, either due to anisotropic
wavefunctions or due to the anisotropic scatterers. We note that the $s$-$d$
model which is sometimes invoked to qualitatively explain the AMR in
ferromagnetic transition metals, is a variation of the anisotropic
wavefunction mechanism. Because of the competing effect of the SOI and
magnetization (which are both present only in the low-mobility $d$ states) one
may expect weak AMR. 

On the other hand, we have shown that (Ga,Mn)As constitutes a textbook example
of a system where strong AMR can be expected since the two agents are mostly
separated: polarized Mn ions act as anisotropic scatterers while the
current-carrying valence-band states bring in the SOI. Quantitatively, the
latter are also partly polarized but we have demonstrated that for the typical
experimental range of Mn dopings in metallic (Ga,Mn)As samples ($x=2\sim
10\%$), the anisotropic scatterer mechanism is dominant. A simple model which
neglects the other two mechanisms provides analytical results which predict
the correct sign of the AMR (resistivity parallel to magnetization is smaller
than perpendicular to magnetization) and identify its origin --- destructive
interference between electric and magnetic part of the scattering potential
(of ionized Mn acceptors) for carriers moving parallel to magnetization.

We gratefully acknowledge Jan Ma\v sek for his advice regarding the range of
validity of the model in Sec.~\ref{sec-IIA}, and the following host of
research-supporting governmental instruments: AV0Z10100521, LC510,
KAN400100652, FON/06/E002 of \hbox{GA \v CR}, and KJB100100802 of \hbox{GA AV}
of the Czech republic, the NAMASTE (FP7 grant No.~214499) and SemiSpinNet
projects (FP7 grant No.~215368), SWAN-NRI,
ONR under Grant No.~onr-n000140610122,
NSF under Grant No.~DMR-0547875, 
and also Pr\ae mium Academi\ae{}.

\bibliographystyle{apsrev}

\end{document}